\documentclass[aps,pre,twocolumn,a4paper,10pt,notitlepage,footinbib,superscriptaddress,longbibliography]{revtex4-1}
\usepackage[english]{babel}
\usepackage{lmodern}
\usepackage[latin9]{inputenc}
\usepackage{endnotes}
\usepackage{amssymb,amsmath,amsfonts}
\usepackage{textcomp}
\usepackage{url}
\usepackage{soul}

\usepackage{graphicx}
\usepackage{dcolumn}
\usepackage{bm}
\usepackage{xcolor}

\begin{document}

\title{Dynamics of polydisperse multiple emulsions in microfluidic channels}
\author{A. Tiribocchi}
\email{Corresponding author: a.tiribocchi@iac.cnr.it, adrianotiribocchi@gmail.com}
\affiliation{Istituto per le Applicazioni del Calcolo CNR, via dei Taurini 19, 00185 Rome, Italy}
\author{A. Montessori}
\affiliation{Istituto per le Applicazioni del Calcolo CNR, via dei Taurini 19, 00185 Rome, Italy}
\author{M. Durve}
\affiliation{Center for Life Nano Science@La Sapienza, Istituto Italiano di Tecnologia, 00161 Roma, Italy}
\author{F. Bonaccorso}
\affiliation{Istituto per le Applicazioni del Calcolo CNR, via dei Taurini 19, 00185 Rome, Italy}
\affiliation{Center for Life Nano Science@La Sapienza, Istituto Italiano di Tecnologia, 00161 Roma, Italy}
\affiliation{Department of Physics and INFN, University of Rome ``Tor Vergata'', Via della Ricerca Scientifica 00133 Rome, Italy.}
\author{M. Lauricella}
\affiliation{Istituto per le Applicazioni del Calcolo CNR, via dei Taurini 19, 00185 Rome, Italy}
\author{S. Succi}
\affiliation{Istituto per le Applicazioni del Calcolo CNR, via dei Taurini 19, 00185 Rome, Italy}
\affiliation{Center for Life Nano Science@La Sapienza, Istituto Italiano di Tecnologia, 00161 Roma, Italy}
\affiliation{Institute for Applied Computational Science, John A. Paulson School of Engineering and Applied Sciences, Harvard University, Cambridge, Massachusetts 02138, USA}

\date{\today}

\begin{abstract}
Multiple emulsions are a class of soft fluid in which small drops are immersed within a larger one and stabilized over long periods of time by a surfactant. We recently showed that, if a {\it monodisperse} multiple emulsion is subject to a pressure-driven flow, a wide variety of non-equilibrium steady states emerges at late times, whose dynamics relies on a complex interplay between hydrodynamic interactions and multi-body collisions among internal drops. In this work we use lattice Boltzmann simulations to study the dynamics of {\it polydisperse} double emulsions driven by a Poiseuille flow within a microfluidic channel. Our results show that their behavior is critically affected by multiple factors, such as initial position, polydispersity index and area fraction occupied within the emulsion. While at low area fraction inner drops may exhibit either a periodic rotational motion (at low polydispersity) or arrange into non-motile configurations (at high polydispersity) located far from each other, at larger values of area fraction they remain in tight contact and move unidirectionally. This decisively conditions their close-range dynamics, quantitatively assessed through a time-efficiency-like factor. Simulations also unveil the key role played by the capsule, whose shape changes can favor the formation of a selected number of non-equilibrium states in which both motile and non-motile configurations are found.
\end{abstract}

\maketitle

\section{Introduction}

Dazzling examples of hierarchical soft materials, i.e. organized states of matter made of multilength scale units \cite{chem}, are ubiquitous in nature. They include bones \cite{wegst_nat}, dna \cite{mare_book}, and cells \cite{bray} as well as liquid crystals \cite{degennes},  mesoporous materials \cite{li_natrev}, bijels \cite{strat_sci} and foams \cite{durian_prl,durian_pre}, to name but a few. Of particular relevance to us are multiple emulsions, a pristine compartmentalized soft fluid made of distinct immiscible drops (often termed as cores, of size up to $100\mu$m) encapsulated within  a larger one and stabilized through a surfactant adsorbed onto their interface \cite{utada2005monodisperse,abate2009,datta2014,ding2019,vladi2017}.  

Due to their fascinating architectural design encompassing various length scales (from the interface of a few nanometers to the diameter of the outer drop of hundreds of micrometers), these materials have found applications in disparate sectors of modern industry, such as pharmaceutics in drug delivery \cite{pays2002,sela_2009,mao2019}, cosmetics in personal care items \cite{lee2001preparation,lee2002effective}, food science in low calories products \cite{Dickins2011,sapei2012,comunian2014,musciol2017} and tissue engineering \cite{rocha2008,chung2012,chan_2013,costantini2014}. Intriguingly, they have recently served as a platform to study cell-cell and cell-bacteria interactions enclosed in a capsule flowing within capillaries \cite{zhang_2013,chan_2013,kaminski2016,choi2016}.  

Multiple emulsions are usually manufactured within microfluidic channels, such as T-junctions and flow-focusing devices \cite{vladi2017},  by means of a single or two-step emulsification process \cite{kim2011one,kim2018,ding2019,clegg2016}, techniques generally guaranteeing a large production rate combined with an ordered design \cite{weitz,abate2009,wang_2011,shum2012}, in which the degree of monodispersity is much higher than that accessible using conventional homogenizers, such as shear mixers \cite{clement2012}. Their mechanical properties can be modulated by properly tuning a number of key physical parameters, such as viscosity of the middle fluid to harden or jell the capsule, surface tension of the drops to modify their shape, amount and type of surfactant solution to prevent coalescence as well as degree of polydispersity \cite{utada2005monodisperse,montessoriprf,montessori_lang}. 
The latter, in particular, can be decisively affected by the hydrodynamic forces mediated by the thin film of fluid separating the cores, whose mechanical stability relies on a delicate balance between surface tension and disjoining pressure \cite{marmottantsoft,raven2009,chan2011}. Such a balance is especially relevant when the emulsion is subjected to an external forcing, such as shear or Poiseuille flow, which is a common situation in microfluidic experiments. Indeed, hydrodynamics may foster collisions among cores when flowing within the microchannel, thus favoring shape deformations which can potentially lead to their breakups and merging \cite{stone,smith,kim2011lab,chen2011}. Such effects would permanently alter the grain size of the cores and ultimately modify their rate of polydispersity, eventually jeopardizing the homogeneity of the material, a feature often required, for example, in the design of soft porous matrices, such as tissue scaffold \cite{costantini2014}. In addition, polydisperse cores may also be produced upstream due to an uncontrolled breakup of the dispersed fluid at the orifice located at the injection channel \cite{sauret2012}. It is thus crucial to understand how hydrodynamics and polydispersity affect the mechanics of a multiple emulsion under flow, in particular the motion of the internal cores and the morphology of the droplets.

In this work, we use lattice Boltzmann (LB) simulations \cite{succi1} to study the dynamics of polydisperse multiple emulsions subject to a Poiseuille flow within a microfluidic channel. A number of relatively recent numerical works have been dedicated to model such physics either considering double emulsions (containing a single core) \cite{coupier_prl,zou2008,abreu2014,wangpre2014,pommella2017,che2018,wang2020,pontrelli2020} or multiple emulsions with distinct monodisperse cores \cite{tao2013,tiribocchi_nat}. In the latter context, we have lately shown that a surprisingly wide variety of non-equilibrium steady states can potentially be found when these systems are subject to a pressure-driven flow \cite{tiribocchi_nat}. They essentially range from states where the cores exhibiting a persistent periodic motion, triggered by a dipolar fluid vortex formed within the shell, to further states in which the cores display a chaotic-like dynamics due to the complex interplay between many-body collisions and hydrodynamic interactions. Here, we take one step forward and show that a wider scenario emerges when considering a polydisperse mixture of drops confined within a fluid capsule. Our multiple emulsion consists of two drops of radii $R_1$ and $R_2$ with $R_2>R_1$ and polydispersity index $h=R_2/R_1$ (see Fig.\ref{fig1}), immersed within larger droplet. Their physics is captured by a multiphase field model \cite{aranson,marenduzzo1}, where the dynamics of a set scalar fields accounting for the density of each droplet obeys a Cahn-Hilliard equation, while that of a vector field representing the fluid velocity is governed by the Navier-Stokes equation. 

Our LB simulations show that the emulsion dynamics depends, in a non-trivial way, on a set of key physical parameters, such as the polydispersity of the cores and their initial position as well as the area fraction $A_f$ they occupy and the morphology of the capsule. At low values of $A_f$, for example, the cores may either exhibit a permanent periodic motion confined within half of the emulsion for long periods of time (for low $h$) or get stuck separately at the front or at the rear of the emulsion (for high $h$), in particular when the large core precedes the small one at the onset of the motion (once the Poiseuille flow is turned on).
At high values of $A_f$,  the cores arrange in non-motile configurations remaining either far from each other or in close contact confined at the leading edge of the capsule,  a behavior generally observed for higher values of $h$ and regardless of their initial position.
This range of dynamic behaviors importantly affects the interaction between the cores,  a quantity evaluated in terms of a dimensionless number gauging the time spent by the drops in close contact during a simulation. Finally, we also show that modifications of the shape of the capsule can foster a range of non-equilibrium steady states partially akin to the ones aforementioned.

The paper is structured as follows. In Section II, we outline the thermodynamic properties of the emulsion and the numerical method. An approximate mapping with physical parameters is also provided. Section III is dedicated to a discussion of the physics at the steady state, in particular the structure of the velocity field and the behavior of the internal cores. This is assessed in terms of their initial position, the area fraction occupied in the emulsion, the polydispersity index and the time spent in close contact.  
We also illustrate the role played by the shell in driving the motion of the cores through changes of its shape. In conclusion, some remarks close the paper.

\section{Method}

\subsection{Equations of motion}

As in previous works \cite{marenduzzo1,tiribocchi_pof,tiribocchi_nat,tiribocchi_prf,tiribocchi_pof2}, we use a multiphase field approach fully incorporating hydrodynamic interactions to model the physics of a multi-core emulsion. Basically, a set of scalar fields $\phi_i({\bf r},t)$, $i=1,....,N$ (where $N$ is the total number of droplets) accounts for the density of each droplet at position ${\bf r}$ and time $t$ while a vector field ${\bf v}({\bf r},t)$ describes the fluid velocity.

Each phase field $\phi_i$ obeys an advection-diffusion equation
\begin{equation}\label{CH_eqn}
 D_t\phi_i=M\nabla^2\mu_i,
\end{equation}
where $D_t=\frac{\partial}{\partial t}+{\bf v}\cdot\nabla$ is the material derivative, $M$ is the mobility and $\mu_i$ is the chemical potential of the $i$-th droplet.

The time evolution of fluid velocity ${\bf v}({\bf r},t)$ is governed by the Navier-Stokes equation which, in the incompressible limit, reads
\begin{equation}\label{NAV_eqn}
  \rho\left(\frac{\partial}{\partial t}+{\bf v}\cdot\nabla\right){\bf v}=-\nabla p +\eta\nabla^2{\bf v}-\sum_i\phi_i\nabla\mu_i.
\end{equation}
Here $p$ is the hydrodynamic pressure and $\eta$ is the dynamic viscosity.

Finally, the chemical potential appearing in Equations (\ref{CH_eqn}) and (\ref{NAV_eqn}) is defined as $\mu_i\equiv\frac{\delta {\cal F}}{\delta\phi_i}$, where ${\cal F}=\int fd^3{\bf r}$ is the free energy
encoding the equilibrium properties of the mixture \cite{degroot,landau}. Its density $f$ is given by
\begin{equation}\label{freeE}
f=\sum_i^N\left(\frac{a}{4}\phi_i^2(\phi_i-\phi_0)^2+\frac{k}{2}(\nabla\phi_i)^2\right)+\sum_{i,j,i<j}\epsilon\phi_i\phi_j,
\end{equation}
where the first two terms ensure the existence of two minima,  $\phi_i=\phi_0$ (with $\phi_0\simeq 2$) inside the $i$th droplet and $\phi_i=0$ outside, separated by a fluid interface.
These two contributions also determine the surface tension $\sigma=\sqrt{8ak/9}$ and the interfacial thickness $\xi=2\sqrt{2k/a}$ \cite{kruger,cates1}, in which $a$ and $k$ are positive constants.
Finally, the last term of Eq.(\ref{freeE}) represents a repulsive potential of strength $\epsilon$, mimicking the effects produced, at the mesoscale level, by a surfactant adsorbed onto the droplet interfaces.

\subsection{Numerical aspects}

Equations (\ref{CH_eqn}) and (\ref{NAV_eqn}) are numerically solved by means of a hybrid LB method \cite{succi1,kruger,sukop,montessori}, in which the advection-relaxation equations are integrated via a finite difference Euler algorithm while the Navier-Stokes equation through a predictor-corrector LB scheme \cite{tiribocchi,tiribocchi4,tiribocchi_soft, tiribocchi_soft2}.

Simulations are run on two dimensional rectangular lattices of size $L_y=800$ (horizontal) and $L_z=170$ (vertical). Periodic boundary conditions are set along the $y$-direction while two parallel flat walls are placed at $z=0$ and $z=L_z$. Here no-slip conditions hold for the velocity field ${\bf v}$, meaning that $v_z(z=0,z=L_z)=0$, while neutral wetting holds for the fields $\phi_i$. The latter ensures that mass flux through the walls is absent (i.e. ${\bf n}\cdot\nabla \mu_i|_{z=0,z=L_z}=0$, where ${\bf n}$ is a unit vector normal the boundaries), and interfaces are perpendicular to the walls (i. e. ${\bf n}\cdot\nabla(\nabla^2\phi_i)|_{z=0,z=L_z}=0$).

\begin{figure}[htp]
\includegraphics[width=1.\linewidth]{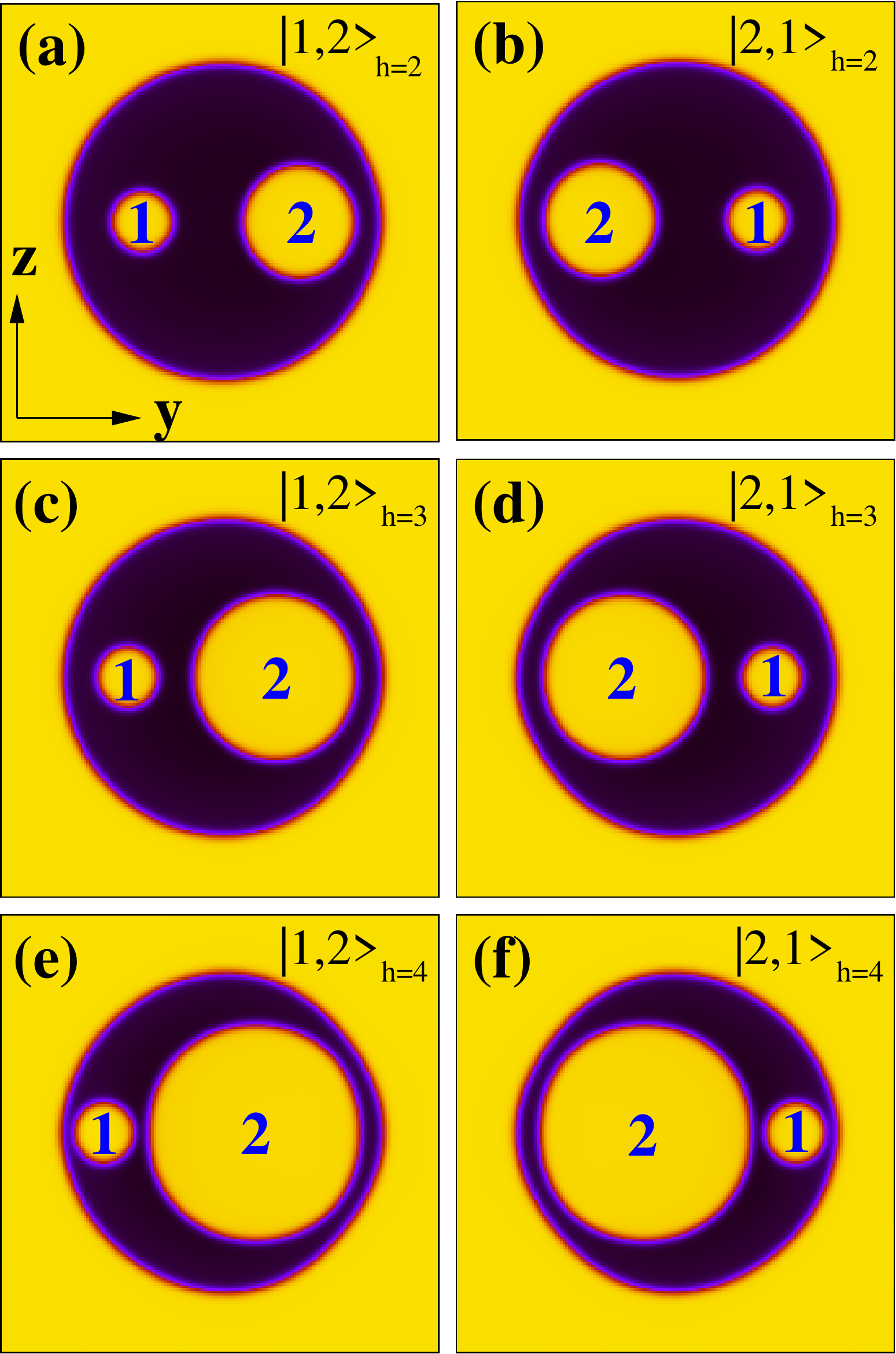}
\caption{Equilibrium configurations of three double emulsions with different sizes of the inner drops, (a-b) $R_1=10$, $R_2=20$ ($h=2$, $A_f\simeq 0.14$); (c-d) $R_1=10$, $R_2=30$ ($h=3$, $A_f\simeq 0.28$) and (e-f) $R_1=10$, $R_2=40$ ($h=4$, $A_f\simeq 0.47$). Only a portion of the microchannel is shown. Symbol $|i,j\rangle_{h=n}$ (where $i,j,n$ are integer numbers) indicates a state in which two cores (such as $1$ and $2$) of different size are positioned next to each other, while $h$ is the polydispersity index.
Snapshots with equal values of $h$ (such as a and b, c and d, e and f) only differ for the initial position of the cores. The radius of the external droplet has been kept fixed to $R_{e}=60$. Colors represent the values of the order parameter $\phi$, ranging from $0$ (black) to $\simeq 2$ (yellow). This applies to all figures.}
\label{fig1} 
\end{figure}

In Fig.\ref{fig1} we show three examples of double emulsions with different polydispersity  index $h=R_2/R_1$, where $R_1$ and $R_2$ are the radii of cores $1$ (small) and $2$ (large), respectively. Emulsions with equal values of $h$ (such as a-b, c-d and e-f) are prepared in two initial states, differing in the initial position of the cores. In Fig.\ref{fig1}a, for example, the core $1$ with radius $R_1=10$ lattice sites is located on the left side of core $2$ having radius $R_2=20$, while the radius of the external droplet is $R_e=60$. This value, kept fixed in the simulations, ensures that contacts between walls and emulsion are minimized. In Fig.\ref{fig1}b the small core is positioned on the right side of the large one. Hence in both cases one has $h=2$ and area fraction occupied by the cores equal to $A_f=\frac{\pi\sum_iR_i^2}{\pi R_{e}^2}\simeq 0.14$. Such states are indicated as $|1,2\rangle_{h=2}$ (Fig.\ref{fig1}a) and $|2,1\rangle_{h=2}$ (Fig.\ref{fig1}b).
In Fig.\ref{fig1}c-d one has $R_1=10$ and $R_2=30$ ($h=3$, $A_f\simeq 0.28$), and in Fig.\ref{fig1}e-f $R_1=10$ and $R_2=40$ ($h=4$, $A_f\simeq 0.47$). Since three drops are considered (two cores and the shell), one needs three phase fields. In particular the fields $\phi_1$ and $\phi_2$ refer to droplets $1$ and $2$, and are positive (equal to $\simeq 2$) within each drop and zero everywhere else. The field $\phi_3$ of the external droplet is positive outside and zero elsewhere.
Once the droplets are relaxed towards a near-equilibrium state, a pressure gradient $\Delta p$ is applied to produce a Poiseuille flow. This is modeled through the inclusion of a body force pushing the emulsion rightwards, along the positive $y$ axis.

If not stated otherwise, following previous works \cite{tiribocchi_nat} the thermodynamic parameters have been chosen as follows: $a=0.07$, $k=0.1$, $M=0.1$, $\eta=1.67$, $0.01<A_f< 0.5$, $\Delta x=1$ (mesh step), $\Delta t=1$ (time step) and $\epsilon=0.05$. In particular, the values of the parameters $a$ and $k$ fix the surface tension $\sigma \simeq 0.08$ and the interfacial width $\xi\simeq 3.5$, while the diffusion constant is $D=Ma=0.007$. Finally, the value of $\epsilon$ is sufficiently high to prevent droplet merging.

A mapping between these simulation parameters and real values can be built by fixing typical length, time and force scale as $L=1\mu m$, $T=10\mu s$ and $F=10nN$ (in simulation units these scales are all equal to one). Hence  this corresponds to a microfluidic channel of length $\sim 1mm$ in which droplets, of diameter ranging from $\sim 10\mu m$ (cores) to $\sim 100\mu m$ (shell), have a surface tension of $1mN/m$ and are immersed in a fluid of viscosity $\eta\simeq 10^{-1}$Pa s (equal viscosity is assumed for the fluid inside and outside the droplets). The Reynolds number, defined as $Re=\frac{\rho D_ev_{max}}{\eta}$ where $v_{max}$ is the maximum speed in the channel and $D_e$ is the diameter of the shell, varies roughly between $0.5$ ($v_{max}\simeq 0.007$, $\Delta p=3\times 10^{-4}$) and  $5$ ($v_{max}\simeq 0.025$, $\Delta p=10^{-3}$), while the capillary number $Ca=\frac{v_{max}\eta}{\sigma}$ ranges from $0.1$ to $1$.

With these numbers, gravity effects can be neglected since, assuming $\Delta \rho/\rho_w=(\rho_w-\rho_o)/\rho_w\sim 0.1$ (where the water density $\rho_w=10^3Kg/m^3$ and a typical oil density $\rho_o\sim 9\times 10^2Kg/m^3$), one has a Bond number $Bo=\Delta\rho g R^2/\sigma\sim 10^{-2}\div 10^{-3}$, where $g$ is the gravity acceleration.

\section{Results}

We start by describing steady states and shape deformations observed when a two core emulsion, with a different polydispersity ratio, is subject to a Poiseuille flow. Afterwards we focus on the contact dynamics among cores observed at different values of $A_f$, and finally we elucidate the role played by the shape of the shell in selecting specific steady states.

\subsection{Steady states shapes and core dynamics}

Starting from the set of equilibrated configurations shown in Fig.\ref{fig1}, we impose a Poiseuille flow 
which pushes the multi-core emulsion along the positive $y$ axis, i.e. the longitudinal direction of the microchannel. Once the flow is switched on, the shell progressively stretches while the internal cores are dragged towards its front. In these conditions, the fluid velocity exhibits two well-defined counter-rotating vortices within the emulsion significantly affecting the dynamics of the internal cores \cite{tiribocchi_nat}. 

\begin{figure*}[htp]
\includegraphics[width=0.5\linewidth]{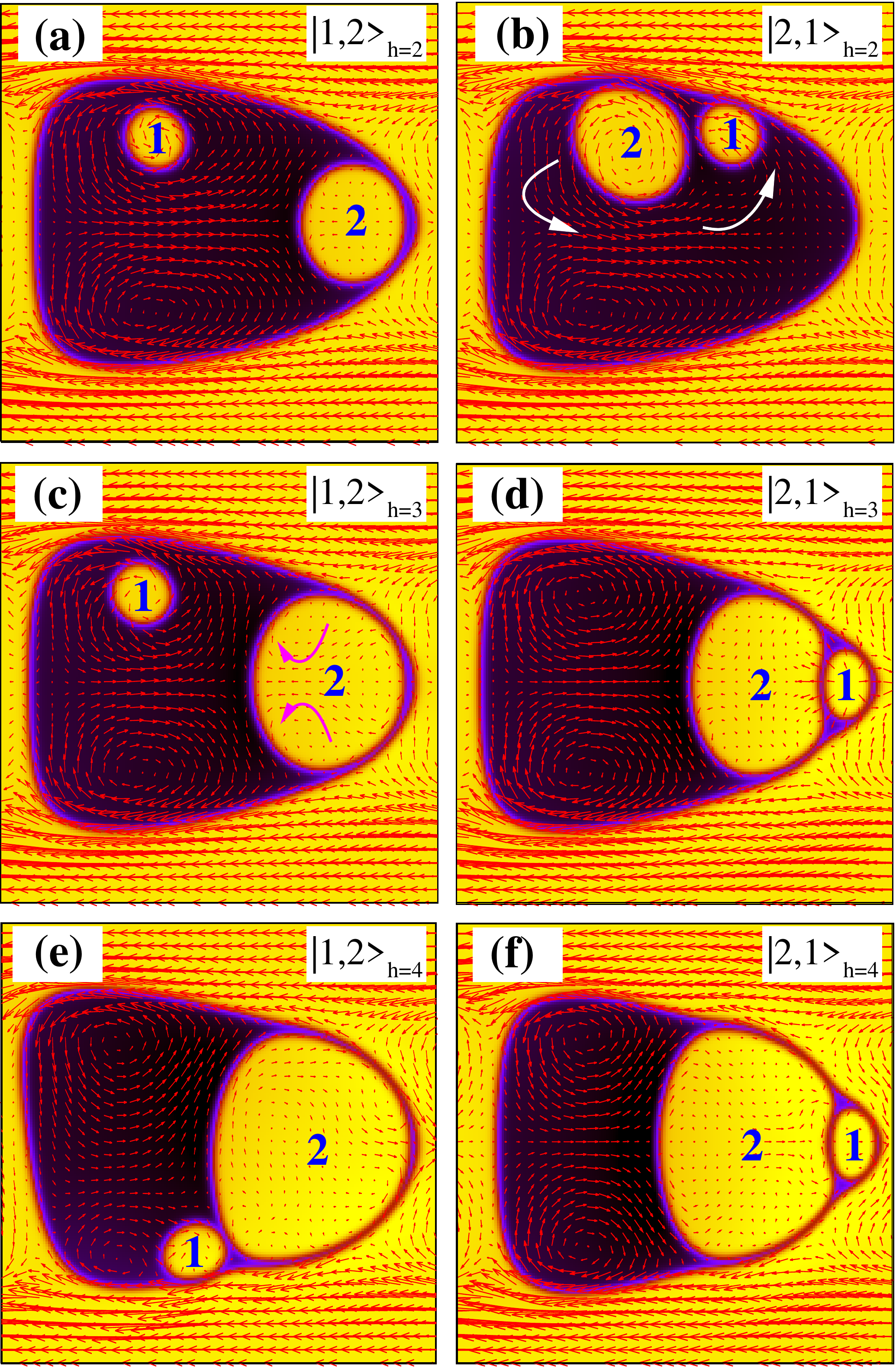}
\caption{Non-equilibrium steady states observed in polydisperse two-core emulsions subject to a Poiseuille flow. Here $Re\simeq 2$ and $Ca\simeq 0.53$. The left column represents states whose initial conditions shown in Fig.\ref{fig1}a-c-e are of the form $|1,2\rangle_{h=2,3,4}$, while the right column stems from initial configurations of the form $|2,1\rangle_{h=2,3,4}$. Red arrows indicate the velocity field computed in the frame of reference of the external droplet. If $h=2$ (a-b) one can identify two regimes, (i) one in which the large core gets stuck at the leading edge of the emulsion while the small one is captured by the upper fluid vortex, and (ii) a further one in which the cores exhibit a periodic motion triggered by the fluid vortex. In both cases the whole emulsion is dragged rightwards by the flow.  White arrows indicate the direction of motion of the cores.  If $h=3$ (c,d) and $h=4$ (e-f), internal drops attain a non-motile configuration, either akin to that shown in (a) or consisting of two cores stuck at the front of the emulsion. Magenta arrows in (c) represent the direction of two counterclockwise vortices formed within the larger core.}
\label{fig2} 
\end{figure*}
If $h=2$, for example, one may observe two distinct scenarios, whose evolution crucially depends on the way the emulsion is initially prepared. 
In the first one, observed starting from the state $|1,2\rangle_{h=2}$, both cores accumulate at the leading edge of the shell and only temporarily arrange in a row. Indeed, such a configuration is unstable to weak perturbations of the flow, an effect due to the coupling between the velocity field and the fluid interfaces \cite{tiribocchi_nat}. This leads to a state in which the large core remains stuck at the front of the emulsion while the small one initially shifts upwards and then is  dragged backward until its motion relative to the shell ceases (see movie M1 \cite{suppl}, Fig.\ref{fig2}a). Following an analogous mechanism,
the small core may alternatively shift downwards remaining confined within the lower sector of the emulsion, a condition achieved by slightly diminishing Reynolds and capillary numbers \cite{tiribocchi_nat}. 
In both cases, at late times the cores and the shell move unidirectionally rightwards and no further changes are observed in their trajectories (see Fig.\ref{fig3}a-b, where the displacement $\Delta {\bf r}_{cm}$ of the cores with respect to the shell is plotted over time). In the second scenario obtained starting from the state $|2,1\rangle_{h=2}$, once the row-like arrangement of the cores at the front of the shell is destabilized, the large core is initially driven backwards by the upper vortex but, unlike the previous case, it is followed by the small one. They move together backwards and then forward, exhibiting a persistent periodic motion along an almost circular trajectory confined within the upper region of the emulsion (see movie M2 \cite{suppl}, Fig.\ref{fig2}b and Fig.\ref{fig3}a-b).  Here too, decreasing $Re$ and $Ca$ leads to a state where both cores display the periodic motion within the lower part of the emulsion. {We incidentally note  that one may also initially align the centers of mass of the cores along the $z$ direction rather than the mid-line. If such conditions are set, under Poiseuille flow the cores are essentially driven forward by the fluid, remaining confined within the regions (upper or lower) selected from scratch.
These results  are in agreement with the ones discussed in Ref.\cite{tiribocchi_nat}, where the dynamics of a {\it monodisperse} multi-core emulsion under Poiseuille flow has been studied.

\begin{figure}[htp]
\includegraphics[width=1.\linewidth]{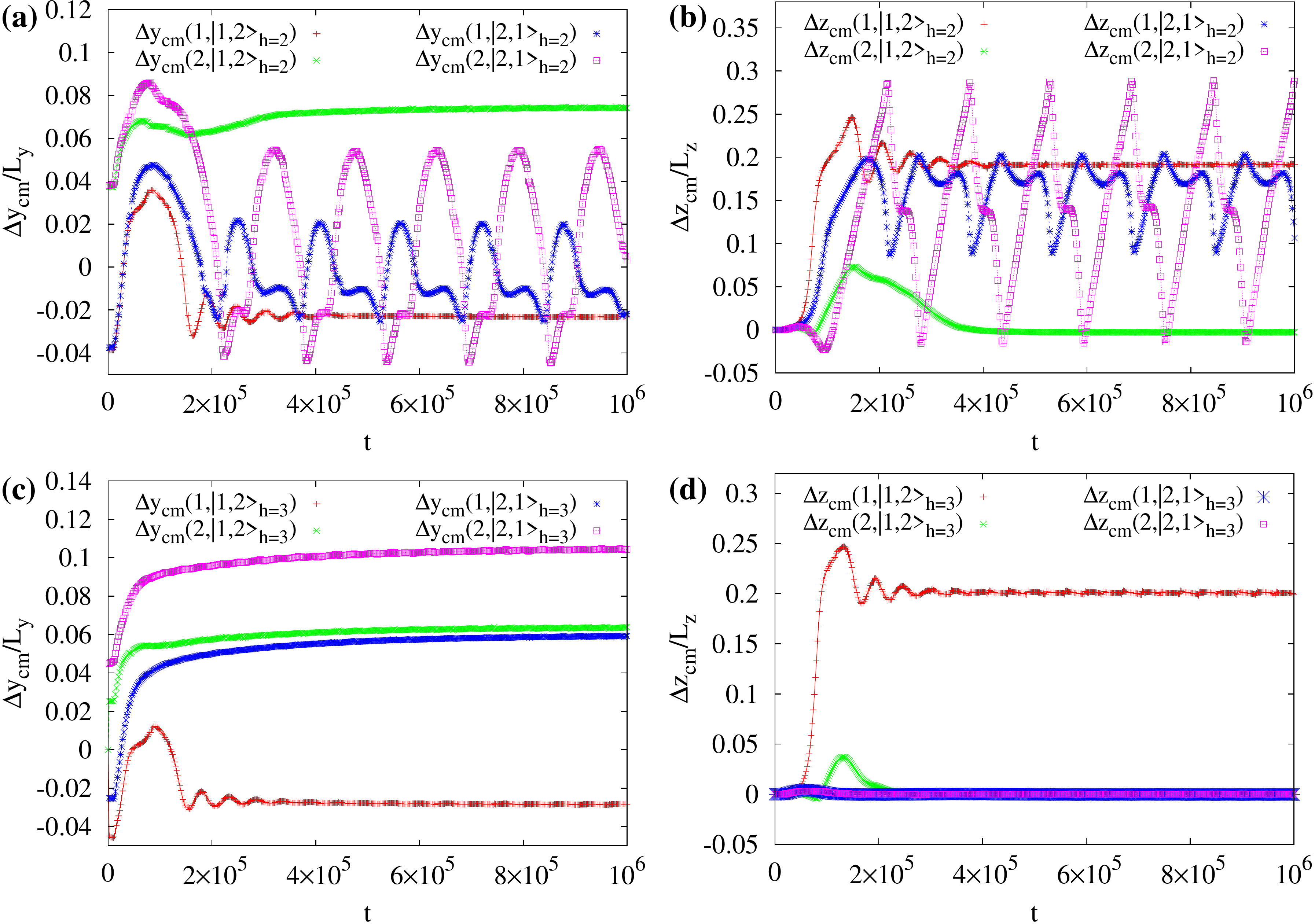}
\caption{Time evolution of the displacement $\Delta y_{cm}$ (left column) and $\Delta z_{cm}$ (right column) of the internal cores. Top panels refers to states of $|1,2\rangle$ and $|2,1\rangle$ with $h=2$, while bottom ones concern those with $h=3$.  If $h=2$ cores may either get transported by the fluid and move unidirectionally or display a persistent periodic motion confined within a region of the emulsion and  triggered by the fluid vortex. If $h=3$, the internal cores are generally dragged by the flow unidirectionally, either remaining sufficiently far from each other or accumulating at the front of the emulsion.}
\label{fig3} 
\end{figure}
If $h=3$ the dynamic behavior is overall simpler although, once again, it is importantly affected by the initial arrangement of the cores. We basically find that, starting from the state $|1,2\rangle_{h=3}$, at late times the drops attain a steady configuration (see Fig.\ref{fig2}c and movie M3 \cite{suppl}) akin to that observed in Fig.\ref{fig2}a whereas, starting from the state $|2,1\rangle_{h=3}$, both cores accumulate at the front of the emulsion where they stick (see Fig.\ref{fig2}d and movie M4 \cite{suppl}). The latter occurs because the large core initially follows the small one once the Poiseuille flow is turned on. Then it approaches and entraps core $1$ at the front of the emulsion, in a region where the effect of the fluid vortices is too weak to destabilize the suspension. In both cases (c-d), at the steady state the cores move unidirectionally at constant speed  without displaying any considerable deviation from their path (see also Fig.\ref{fig3}c-d). Note incidentally that besides the two fluid recirculations formed within the shell (the black region), two further vortices appear in the large core (highlighted by magenta arrows), counter-rotating with respect to the contiguous one due to the continuity of the field at the interface. Such structures look seemingly  absent in smaller drops (such as core $1$ in Fig.\ref{fig2}d), very likely because of a lack of sufficient resolution of the present mesoscale approach, which captures with very good accuracy fluid vortices whose typical size is generally comparable with that of the pertaining droplet (which is of the order of a few microns).

At increasing values $h$, such as $h=4$ (see Fig.\ref{fig2}e-f), the steady states show analogous dynamic features to the ones observed for $h=3$, except for the shape attained by the shell. Indeed,  while at low values of $h$ the external interface attains a rather well-defined  bullet-shape profile (such as in Fig.\ref{fig2}a-b), for higher values it develops permanent local bulges at the front due to the close contact with the internal cores. The shape may, for instance, stretch along the flow direction (see Fig.\ref{fig2}d and f) or acquire a protuberance located at the top rear (see Fig.\ref{fig2}e), resulting from the asymmetry of the vortices induced by the small drop stuck in the lower region. The dynamics of the cores discussed in Fig.\ref{fig2} qualitatively holds for decreasing values of $Re$ and $Ca$ as well. Setting $Re\simeq 0.5$, $Ca\simeq 0.1$ and $h=2$, for example, at the steady state the shell acquires a more rounded shape, a condition allowing the cores to escape from its front and migrate either downwards or upwards where they remain confined, exhibiting a dynamics akin to that shown in Fig.\ref{fig2}b.

These results suggest that changing the polydispersity index could potentially suppress the internal motion of the cores (thus stabilize the emulsion) and significantly affect their close contact dynamics, while only mildly alter the shape of the emulsion, which basically remains that of a projectile usually observed for droplets subject to a Poiseuille flow \cite{tiribocchi_nat}.
The next section will be dedicated to a quantitative assessment the interaction of cores in terms of the time they spend in close contact.

\subsection{Close-contact dynamics among cores}

In a previous work \cite{tiribocchi_pof2} we showed that, when the polydispersity index increases, contacts among cores are favoured in multiple emulsions under shear flow. Assessing interactions and reciprocal distance may be relevant, for example, in microbiological experiments in which cells  dispersed in an aqueous environment (like our cores encapsulated within the emulsion) come in contact with pathogenic bacteria causing human disease \cite{kaminski2016}.

The close-contact dynamics can be approximately evaluated by comparing the reciprocal distance $|\Delta {\bf r}_{cm}|=|{\bf r}_{cm,1}-{\bf r}_{cm,2}|$ between the centers of mass of each core with the distance $d = R_1 + R_2 + l$, where $R_1$ and $R_2$ are the radii of the cores and $l$ is the length of the thin film separating opposite interfaces in  contact.  Assuming a width of interface $\xi$ approximately equal to $4 \div 5$ lattice sites, we set $l=2\xi$. If $|\Delta {\bf r}_{cm}|<d$ the cores are considered at a sufficiently close distance to temporarily sustain the film of fluid, whereas if $|\Delta {\bf r}_{cm}|>d$ they are too far away and reciprocal interaction is negligible. This quantity provides a reasonable measurement of the interaction as long as the shape deformations of the cores do not significantly depart from a near-spherical one. 

\begin{figure}[htp]
\includegraphics[width=1.\linewidth]{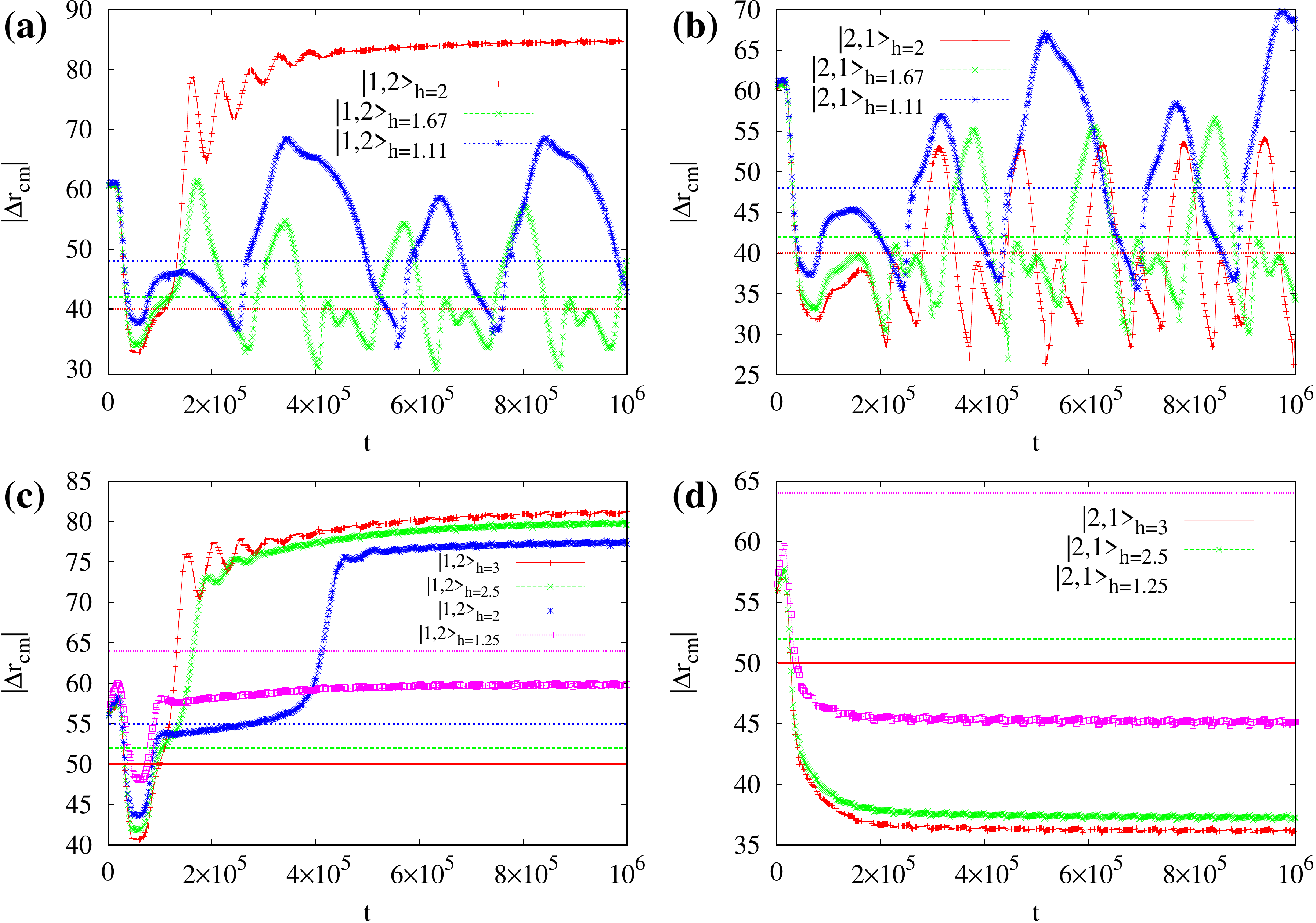}
\caption{Time evolution of the distance $|\Delta {\bf r}_{cm}|$ between the centers of mass of the cores, starting their motion from states of the type $|1,2\rangle$ (a-c) and $|2,1\rangle$ (b-d) and for different values of $h$. Horizontal lines indicate the distance $d =R_1+R_2+l$ below which cores interact, since opposite interfaces belonging to different drops sustain a temporary film of fluid. Assuming an interface width $\xi$ approximately equal to $4-5$ lattice sites, one has $l=2\xi$.  Thus  in (a) and (b) $d=40$ ($R_1=10$, $R_2=20$, $l=10$) for $h=2$ (red horizontal line), $d=42$ ($R_1=12$, $R_2=20$, $l=10$) for $h=1.67$ (green horizontal line) and $d=48$ ($R_1=18$, $R_2=20$, $l=10$) for $h=1.11$ (blue horizontal line). Note that here $R_2$ is fixed at $20$ and $R_1$ increases. In (c) and (d) $d=50$ ($R_1=10$, $R_2=30$, $l=10$) for $h=3$ (red horizontal line), $d=52$ ($R_1=12$, $R_2=30$, $l=10$) for $h=2.5$ (green horizontal line), $d=55$ ($R_1=15$, $R_2=30$, $l=10$) for $h=2$ (blue horizontal line), $d=64$ ($R_1=24$, $R_2=30$, $l=10$) for $h=1.25$ (purple horizontal line). Here $R_2$ is fixed at $30$ and $R_1$ increases.}
\label{fig4} 
\end{figure}

\subsubsection{Dynamics at low values of $A_f$}

In Fig.\ref{fig4}a-b we show the time evolution of $|\Delta {\bf r}_{cm}|$ for a multiple emulsion prepared either in an initial state of type $|1,2\rangle$ (in which the smaller drop $1$ is located on the left side of the larger one $2$) or in a state of type $|2,1\rangle$ (the other way round), with area fraction $A_f$ occupied by the cores ranging between $0.14$ ($h=2$) and $0.2$ ($h=1.11$). If, for example, $h=2$ ($R_1=10$ and $R_2=20$) one has $d=40$ and  $|\Delta {\bf r}_{cm}|$ either attains a constant value higher than $d$ (i.e. the cores remain sufficiently far from each other and collisions are negligible, red/pluses plot of Fig.\ref{fig4}a) or exhibits an oscillating behavior intersecting the horizontal line at $d=40$ multiple times (red/pluses plot of Fig\ref{fig4}b). The latter behavior essentially means that inner cores periodically come close ($\Delta {\bf r}_{cm}<d$) and detach ($\Delta {\bf r}_{cm}>d$) from early times on.  For decreasing values of $h$, the size of the cores approaches the monodisperse limit, a condition achieved by augmenting the radius of the smaller drop $1$ while keeping constant that of the larger one $2$. In these cases ($h=1.67$ and $h=1.11$, green/crosses and blue/asterisks respectively), once again the cores display a persistent circular motion triggered by the fluid vortex, and come in close contact multiple times. Their reciprocal distance shows a rather regular oscillating pattern once the steady state is attained (at approximately $t=2\times 10^5$ time-steps), remaining below the interaction distance for long periods of time. In the next section we will show that, unlike the physics just discussed, at higher values of $A_f$ the contact dynamics is significantly different, despite the values of $h$ will be akin to the ones considered so far.

\subsubsection{Dynamics at intermediate values of $A_f$}

In Fig.\ref{fig4}c-d we show the time evolution of $|\Delta {\bf r}_{cm}|$ for a multiple emulsion prepared either in a state $|1,2\rangle$ (c) or $|2,1\rangle$ (d) in which $A_f$ ranges from $0.28$ ($h=3$, $R_1=10$ and $R_2=30$) to $0.41$ ($h=1.25$, $R_1=24$ and $R_2=30$). If $h=3$ one has $d=50$ and two opposite scenario occurs. Starting from a $|1,2\rangle$ state, for instance, the cores come into contact only at early times while afterwards detach and remain separate (Fig.\ref{fig4}c, red/pluses plot), with the large drop stuck at the leading edge of the emulsion and the small one in its bulk. In stark contrast, starting from a $|2,1\rangle$ state, the two cores soon stick together and $|\Delta {\bf r}_{cm}|$ falls well below $d$ for the entire simulation (Fig.\ref{fig4}d, red/pluses plot). 

Note in particular that, unlike the cases discussed at low values of $A_f$, here for decreasing values of $h$ the oscillating dynamics is replaced by a fully unidirectional motion, in which either the cores remain far apart for a large portion of time (Fig.\ref{fig4}c, $h=2.5$ and $h=2$) or they stay in very close contact all the time (Fig.\ref{fig4}c $h=1.25$ and Fig.\ref{fig4}d $h=2.5$ and $h=1.25$). In Fig.\ref{fig4}d, in particular, $|\Delta {\bf r}_{cm}|$ at late times is considerably below the corresponding value of $d$, an indication that inner drops are squeezed by the flow, and the distance between the centers of mass significantly shortens.  

\subsubsection{Time efficiency factor}

A more quantitative estimate of the effect of the polydispersity on the close contact dynamics can be gauged by introducing an efficiency-like parameter $\Lambda=T_{int}/T_{tot}$, defined as the ratio between the time $T_{int}$ spent by the cores in close proximity (i.e. when $|\Delta {\bf r}_{cm}|<d$) and the total observation time $T_{tot}$ (i.e. the simulation time, corresponding to $10^6$ time steps). In Fig.\ref{fig5} we show how $\Lambda$ varies with $h$, for double emulsions initially prepared in the state $|1,2\rangle$ (red/pluses and blue/asterisks) and in the state $|2,1\rangle$ (green/crosses and magenta/squares). In all cases, $h$ is varied by changing the radius of the small drop (say $1$) and keeping fixed that of the large one (say $2$). Thus, for example, the plot $|1,2\rangle$ type (a) is obtained by starting from a configuration like the one shown in Fig.\ref{fig1}a and by either decreasing or increasing the radius of drop $1$. This changes $h$ from $1.25$ ($R_1=16$, $R_2=20$) to $2.5$ ($R_1=8$, $R_2=20$). An analogous  scheme is used for the other plots. 

Our results prove, once again, that $\Lambda$ is critically affected by i) initial position of the cores and ii) area fraction they occupy. Indeed, considering the states $|1,2\rangle$, $\Lambda$ attains either intermediate or high values at low $h$ (lower than $2$), since the cores either remain stick together or exhibit an approximate circular motion in which they periodically approach and drift apart. At high $h$ (larger than $2$), $\Lambda$ drastically reduces since the cores stay far away from each other. On the contrary, emulsions starting from a $|2,1\rangle$ state generally show high values of $\Lambda$ for all values of $h$ explored, since either cores move periodically along circular path or squeeze together at the leading edge of the external interface. Note also that $\Lambda$ is generally equal to or higher than $0.5$ for low $h$, namely when the area fraction $A_f$ becomes larger than $\simeq 0.15$ (i.e. the size of the small drop approaches that of the large one), whereas $\Lambda$ can decrease for higher $h$, since $A_f$ diminishes and cores are more easily advected by the velocity field, thus they spend a shorter amount of time in contact.

\begin{figure}[htp]
\includegraphics[width=1.\linewidth]{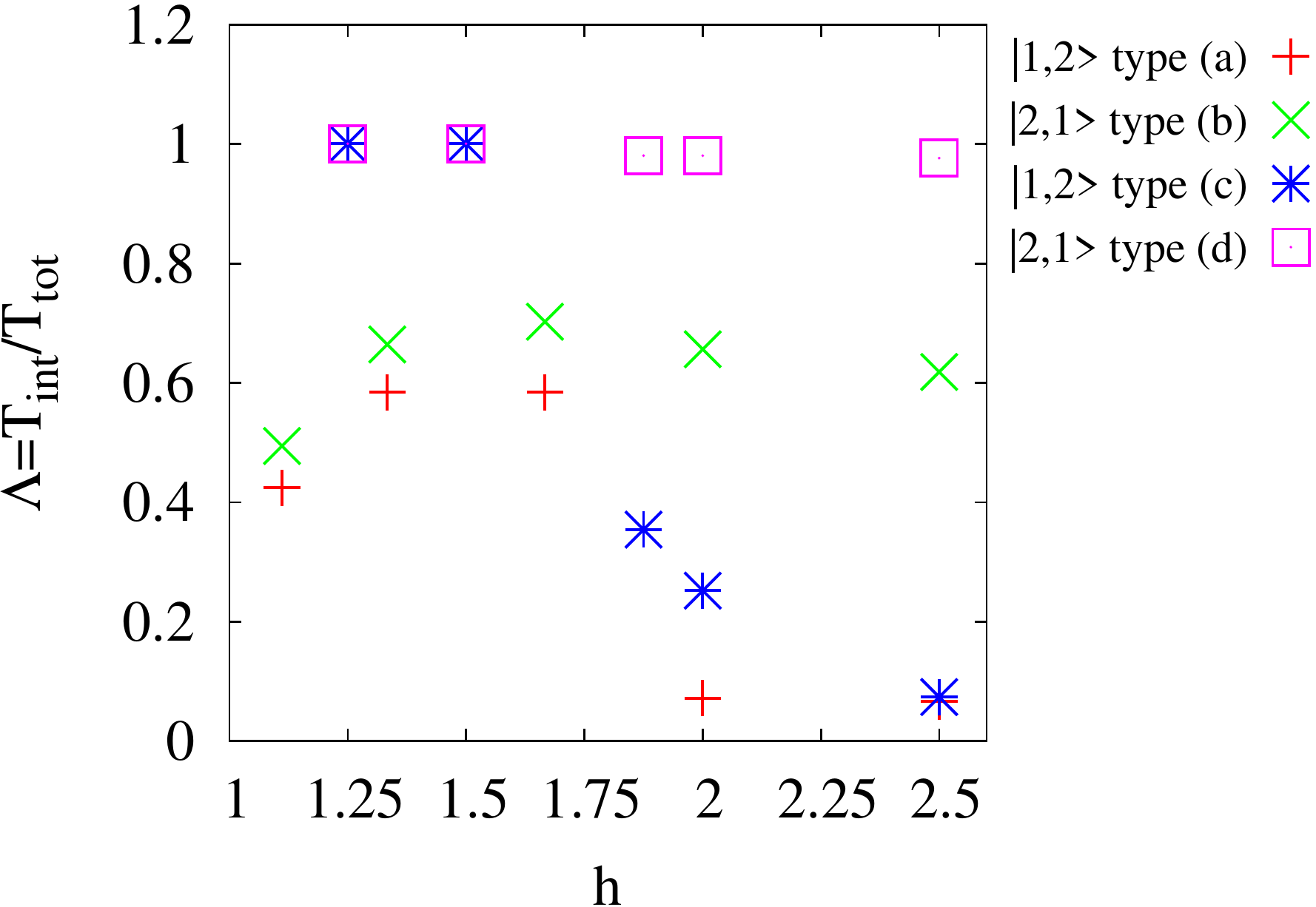}
\caption{Here we plot the ratio $\Lambda=T_{int}/T_{tot}$ for different values of $h$. The quantity $T_{int}$ represents the time (in simulation units) in which two cores are in close contact and sustain a thin film of fluid, an effect occurring when $|\Delta {\bf r}_{cm}|<d$, while $T_{tot}=10^6$ is the total simulation time. Note that in systems starting from a $|1,2\rangle$ configuration (such as those in Fig.\ref{fig1}a-c in which drop $1$ is smaller than drop $2$), the ratio $T_{int}/T_{tot}$ is high for low values of $h$ (i.e. when drops have a comparable size and occupy a large portion of the emulsion) and diminishes as $h$ increases (i.e. the size of droplet $1$ lessens while that of drop $2$ remains fixed). On the contrary, systems starting from a $|2,1\rangle$ configuration display an approximately constant and high value of $\Lambda$.}
\label{fig5} 
\end{figure}

It is finally worth observing that the behavior of $\Lambda$ obtained for a multiple emulsion subject to a Couette-like flow (discussed in Ref.\cite{tiribocchi_pof2}) shows a marked difference with that observed in this paper. Indeed, while in the former $\Lambda$ has a well defined trend in which it typically augments for increasing values of $h$ and essentially regardless of the initial conditions, in the presence of a Poiseuille flow the scenario is more complex since, beside the polydispersity,  both initial conditions and area fraction ultimately affect the fate of the contact dynamics. 

\subsection{Shape of the capsule}
Before concluding, we dedicate this last section to pinpointing the effect produced by the external shell on the motion of the internal cores. Indeed, although our results suggest that polydispersity has only a mild effect on the shape of the shell (as long as $A_f$ remains lower than approximately $0.5$, indeed the highest values considered in the present study), one may wonder whether shape modifications of the external interface could favour a predefined dynamic behavior of the cores. Besides a theoretical interest {\it per se}, the characterization of such shape changes holds a significant relevance in practical applications, such as in drug delivery where, for example, the time release of the drug, usually stored within the cores \cite{pays2002}, is expected to occur faster in regions of high shape deformations  \cite{pontrelli2020}. 

\begin{figure*}[htp]
\includegraphics[width=1.\linewidth]{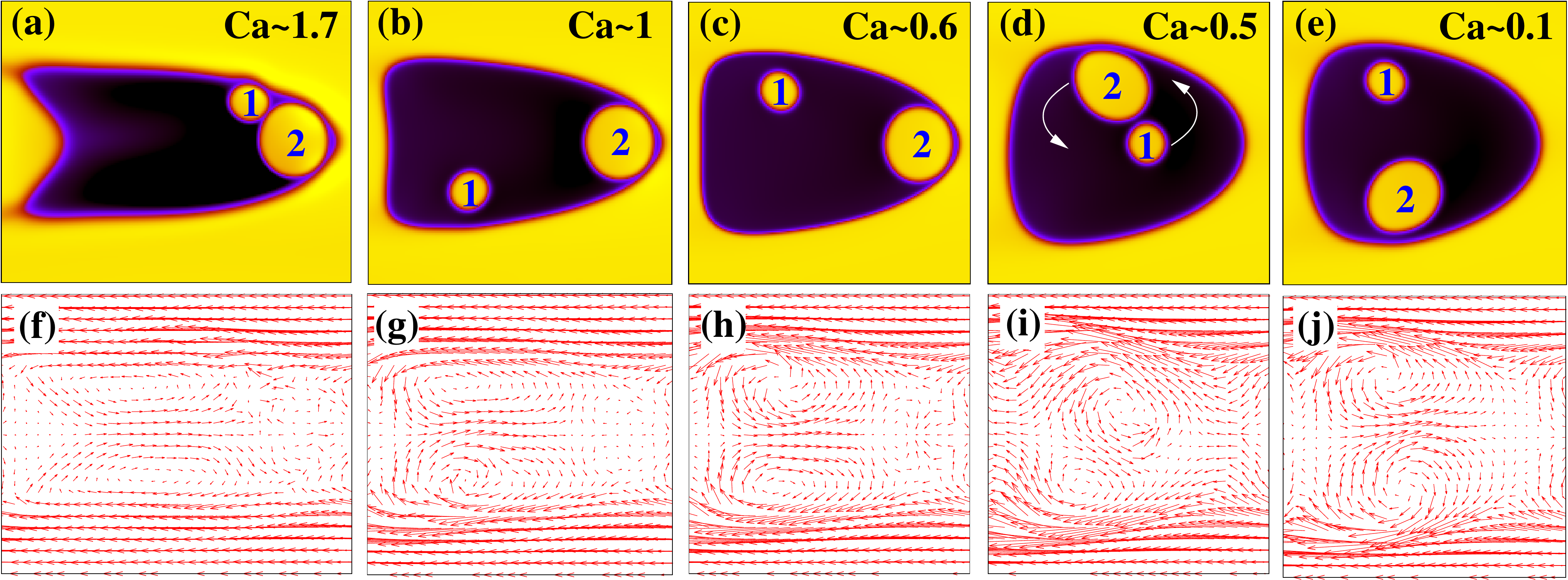}
\caption{The top row shows a selection of nonequilibrium steady states of a two-core emulsion with $h=2$ (initially prepared in the state $|1,2\rangle$) observed by varying the capillary number of the shell while keeping the Reynolds number fixed. At high values of Ca (a) the external droplet exhibits a highly squeezed projectile-like shape which essentially prevents the internal motion of the cores. Decreasing Ca (b-c), the shell acquires a finger-like configuration in which the cores separate and comove with the external droplet. Further diminishing Ca (d-e), the shell takes on a rounded shape in which cores may either rotate periodically or migrate towards separate regions of the emulsion. The bottom row (f-j) shows the velocity field computed in the frame of reference of the shell.}\label{fig6} 
\end{figure*}

In Fig.\ref{fig6} we show a selection of nonequilibrium steady states of a two-core emulsion with $h=2$ initially prepared in a state $|1,2\rangle$. Here $Re$ is kept fixed to $\sim 2.5$ while $Ca$ is varied between $\sim 0.1$ and $\sim 2$ by changing the surface tension $\sigma$ of the external interface. 
Like in previous simulations, the emulsion is driven rightwards by a Poiseuille flow which, besides modifying the shape of the shell, drags the cores along the same direction. Once they approach the front of the external interface, their subsequent dynamic behavior is found to decisively depend on the shape of the emulsion. At high values of $Ca$ (see Fig.\ref{fig6}a and f) the capsule acquires a highly squeezed projectile shape exhibiting a reentrant deformation made of two sharp symmetric bulges located at the rear. Such a structure considerably alters the typical pattern of the velocity field, which now displays two deeply stretched vortices rotating counterclockwise. In this configuration, both cores arrest their internal motion at the leading edge of the emulsion where they remain stuck. This is likely due to a lack of space sufficient to allow for their internal movement and, concurrently, to a "weak" coupling with the velocity field, which, in this state, is unable to trigger their motion. 
At decreasing $Ca$ (see Fig.\ref{fig6}b-c and g-h), the two rear protuberances disappear and the shell attains a wider finger-like structure. This allows for a temporary internal motion of the small cores, which are located either in upper or in the lower part of the emulsion at the steady state.  Now two well defined counter-rotating vortices form within the shell, though still not capable of generating a net internal motion.
Further diminishing $Ca$ (Fig.\ref{fig6} d-e and i-j), the capsule attains a large rounded shape at the steady state.
Here the two symmetric vortices are wide enough to promote the motion of the cores, which either show a periodic dynamics (d) or remain trapped within two separate regions of the emulsion (e).

These results show that shape changes of the shell considerably modify the dynamics of the internal cores, thus further highlighting the multifaceted structure of the parameter phase space determining the fate of these systems, in stark contrast with the ones of the simple liquids they are made of (such as water and oil).

\section{Conclusions}

To summarize, we have simulated, by using lattice Boltzmann methods, the dynamic behavior of a polydisperse multiple emulsion subject to a Poiseuille flow within a microfluidic channel, using a setup inspired to realistic lab experiments. To elucidate the physics, we have considered simple realizations of such emulsions, made of two cores of different size suspended within a larger drop. Despite the easy design, our results provide evidence of a complex scenario in which the properties of the mixture depend on a number of key features, like initial position and area fraction occupied by the cores as well as polydispersity index and shape of the surrounding shell.

At low area fraction (generally $A_f<0.25$), for example, the cores may either display a persistent periodic motion confined within a region of the emulsion for low $h$ (in agreement with previous studies \cite{tiribocchi_nat}), or arrange into non-motile configurations, located far from each other, for high $h$. This behavior crucially depends on the way the emulsion is initially prepared, basically whether the large core precedes or follows the small one once the Poiseuille flow is turned on. At high area fraction, the cores remain firmly glued together moving unidirectionally with the flow for low $h$, or may also disconnect at high $h$. 

Such dynamics also affects the time spent by the cores in contact, a phenomenon once again  depending, in a non-trivial manner, on the aforementioned parameters. This effect has been assessed in terms of the ability of the cores to sustain a film of fluid formed between opposite interfaces during the motion and is quantitatively evaluated through the parameter $\Lambda$. 
In particular, we find that if the large core initially precedes the small one (state $|1,2\rangle$),  at low values of  $h$ $\Lambda$ ranges from $0.5$ to $1$, i.e.  almost the entire simulation time. On the contrary, increasing 
$h$, $\Lambda$ dramatically decreases since both cores place far apart and their interaction becomes negligible. If the large core follows the small one (state $|2,1\rangle$), $\Lambda$ remains considerably high regardless of the values of $h$, since
the cores either periodically approach and separate or conjoin at the front the external interface.  In addition, our findings suggest that the dynamics of the cores can be controlled by changes of capsule shape, ranging from a highly elongated structure observed at high values of $Ca$ to an approximately circular one at low values of $Ca$. 
Their motion is guided by a typical double counter-rotating vortex  which displays a considerably squeezed pattern at high $Ca$ in contrast to a well-defined rounded motif at low $Ca$. It is worth mentioning that the mechanical properties of the emulsion can be also modified by releasing the approximation of equal viscosity between dispersed, middle and outer fluid adopted in this work. Increasing, for example, the viscosity of cores and shell is expected to harden the suspension and affect the structure of the velocity field \cite{utada2005monodisperse}, thus the dynamics of the cores as well. On the contrary, reducing their viscosity would gel the emulsion, likely favouring the breakup of the drops \cite{Park2012}.

We finally note that the scenario described in this paper offers a perspective wider than that discussed in previous studies \cite{tiribocchi_nat}, in particular regarding the role played by the polydispersity. Notwithstanding, several fundamental issues remain open. Delving, for example, into the physics of a multiple emulsion under flow when the volume fraction of the cores considerably overcomes the close packing fraction limit of hard spheres represents a challenging problem. Indeed, the encapsulation of a high packing fraction of drops requires a high control over a number of crucial parameters, such as the surface tension of the shell, the viscosity of the fluids involved and the concentration of the surfactant solution \cite{weitz,guzowski2015}. This is fundamental to avoid effects potentially compromising the design, like rupture of the capsule (in the worst case scenario) or the merging of the droplets, a phenomenon that would permanently alter the topological properties of the emulsion. Such physics is expected to be relevant, for instance, in a highly packed multiple emulsion crossing a narrow constriction \cite{montessori_lang}, where permanent shape deformations and memory-like effects result from the combined action of confinement and viscous dissipation. Understanding, for example, whether topological transitions (such as T1 events) occur mainly in the bulk or in the periphery of the emulsion is still an open problem, as well as determining whether a yield stress marks a transition from a solid-like to a fluid behavior \cite{sbragaglia3}.

\section*{Acknowledgments}
The authors acknowledge funding from the European Research Council under the European Union's Horizon 2020 Framework Programme (No. FP/2014-2020) ERC Grant Agreement No.739964 (COPMAT).

\bibliography{bibliography}

\end{document}